\documentclass[sigplan,screen, sigconf]{acmart}

\AtBeginDocument{%
  \providecommand\BibTeX{{%
    \normalfont B\kern-0.5em{\scshape i\kern-0.25em b}\kern-0.8em\TeX}}}

\usepackage{enumitem}
\usepackage{scalerel}
\usepackage{graphicx}

\usepackage{listings}
\usepackage{subfig}
\usepackage{xcolor}
\usepackage{multirow}
\usepackage{boldline}
\usepackage{flushend}
\usepackage[flushleft]{threeparttable}
\definecolor{codegreen}{rgb}{0,0.6,0}
\definecolor{codegray}{rgb}{0.5,0.5,0.5}
\definecolor{codepurple}{rgb}{0.58,0,0.82}
\definecolor{backcolour}{rgb}{0.95,0.95,0.92}

\lstdefinestyle{htmlStyle}{
    backgroundcolor=\color{white},
    commentstyle=\color{codegreen},
    keywordstyle=\color{blue},
    numberstyle=\tiny\color{codegray},
    stringstyle=\color{codepurple},
    basicstyle=\footnotesize\ttfamily,
    breakatwhitespace=false,
    breaklines=true,
    captionpos=b,
    keepspaces=true,
    numbers=left,
    numbersep=5pt,
    showspaces=false,
    showstringspaces=false,
    showtabs=false,
    tabsize=2,
    language=HTML,
}
\definecolor{lightblue}{RGB}{204, 212, 244}

\definecolor{custom-blue}{rgb}{0,0,0}

\usepackage{amsthm}

\theoremstyle{definition}
\newtheorem{definition}{Definition}[section]

\theoremstyle{remark}

\setcopyright{acmlicensed}
\acmDOI{10.1145/3650212.3680349}
\acmYear{2024}
\copyrightyear{2024}
\acmISBN{979-8-4007-0612-7/24/09}
\acmConference[ISSTA '24]{Proceedings of the 33rd ACM SIGSOFT International Symposium on Software Testing and Analysis}{September 16--20, 2024}{Vienna, Austria}
\acmBooktitle{Proceedings of the 33rd ACM SIGSOFT International Symposium on Software Testing and Analysis (ISSTA '24), September 16--20, 2024, Vienna, Austria}
\acmSubmissionID{issta24main-p707-p}
\received{2024-04-12}
\received[accepted]{2024-07-03}

\begin{document}

\title{Segment-Based Test Case Prioritization: A Multi-objective Approach}

\author{Hieu Huynh}
\orcid{0000-0003-2310-1376}
\affiliation{%
  \institution{Katalon Inc}
  \city{Ho Chi Minh city}
  \country{Vietnam}
}
\email{hieu.huynh@katalon.com}

\author{Nhu Pham}
\orcid{0009-0009-7643-6250}
\affiliation{%
  \institution{Katalon Inc}
  \city{Ho Chi Minh city}
  \country{Vietnam}
}
\affiliation{%
  \institution{University of Science, VNU-HCM}
  \city{Ho Chi Minh city}
  \country{Vietnam}
}
\email{nhu.pham@katalon.com}

\author{Tien N. Nguyen}
\orcid{0009-0006-7962-6090}
\affiliation{%
  \institution{University of Texas at Dallas}
  \city{Texas}
  \country{USA}
}
\email{tien.n.nguyen@utdallas.edu}

\author{Vu Nguyen}
\orcid{0000-0002-0594-4372}
\affiliation{%
  \institution{Katalon Inc}
  \city{Ho Chi Minh}
  \country{Vietnam}
}
\affiliation{%
  \institution{University of Science}
  \city{Ho Chi Minh city}
  \country{Vietnam}
}

\affiliation{%
  \institution{Vietnam National University}
  \city{Ho Chi Minh city}
  \country{Vietnam}
}
\email{nvu@fit.hcmus.edu.vn}

\renewcommand{\shortauthors}{Huynh et al.}

\begin{abstract}
Regression testing of software is a crucial but time-consuming task, especially in the context of user interface (UI) testing where multiple microservices must be validated simultaneously. Test case prioritization (TCP) is a cost-efficient solution to address this by scheduling test cases in an execution order that maximizes an objective function, generally aimed at increasing the fault detection rate. While several techniques have been proposed for TCP, most rely on source code information which is usually not available for UI testing. In this paper, we introduce a multi-objective optimization approach to prioritize UI test cases, using evolutionary search algorithms and four coverage criteria focusing on web page elements as objectives for the optimization problem. Our method, which does not require source code information, is evaluated using two evolutionary algorithms (AGE-MOEA and NSGA-II) and compared with other TCP methods on a self-collected dataset of 11 test suites. The results show that our approach significantly outperforms other methods in terms of Average Percentage of Faults Detected (APFD) and APFD with Cost (APFDc), achieving the highest scores of 87.8\% and 79.2\%, respectively. We also introduce a new dataset and demonstrate the significant improvement of our approach over existing ones via empirical experiments. The paper’s contributions include the application of web page segmentation in TCP, the construction of a new dataset for UI TCP, and empirical comparisons that demonstrate the improvement of our approach.

\end{abstract}



\maketitle

\section{Introduction}
Regression testing is a process performed each time a new version of the software is released to ensure the quality of the software under test. The test suite is usually reserved for testing later releases as the software evolves. Running these test suites consumes a lot of time and can account for half the cost of software maintenance \cite{rothermel1999test}. As for UI testing, the cost is even higher as a web application is a combination of many services. Each interaction on the web page is involved with multiple services. In order to test the behaviour as well as the interactions between components of the web page, multiple services should be validated in one test and the verdict for the test is failed if one of the microservices is failed and vice versa \cite{yu2019terminator}. For large test suites, testers may want to prioritize some test cases with the highest priority first so that the faults are detected sooner and hence, the bugs can be fixed sooner. In such cases, test case prioritization (TCP) is leveraged. 

Test case prioritization schedules test cases in an execution order that maximizes an objective function \cite{rothermel1999test}. An objective may be chosen accordingly to what testers want to prioritize during testing process. One common objective of TCP is to increase the rate of fault detection, which is equivalent to revealing defects sooner in testing procedure. Thanks to that, developers can identify and fix bugs sooner. 
Initially, TCP may seem less critical for small test suites, as random execution can be convenient and equally effective \cite{rothermel1999test}. However, for time-consuming test suites, TCP proves cost-efficient by prioritizing a subset of test cases over the entire suite while aiming to achieve objectives. TCP gains greater significance in UI testing, where tests often take more time than unit tests. Automated UI testing, being black-box, requires no knowledge of source code or application architecture \cite{yu2019terminator}.

There are several techniques introduced to TCP. Search-based is the most utilized approach. Each permutation of test cases is considered a candidate solution and the optimal solution should be discovered heuristically \cite{li2007search} based on coverage information. Coverage-based \cite{rothermel1999test, bryce2011test, hao2015optimal, mahdieh2020incorporating} and history-based \cite{kim2002history,park2008historical,wang2016history} techniques are also among the most popular approaches and show effectiveness despite being coined long ago. Coverage-based approaches aim to cover target items as many as possible in the hope that a higher coverage would result in a higher probability of fault exposure. However, most coverage-based techniques need source code information (statements, functions, branches) while all this information is usually not available for UI testing. History-based approaches use the verdicts from past cycles and need many runs to yield good results \cite{nguyen2021rltcp}. Similarity-based approaches are another set of methods that demonstrate their effectiveness \cite{fang2014similarity, miranda2018fast}. These approaches focus on prioritizing test cases that differ from the already prioritized ones, emphasizing the diversity of test cases. Yu et al. \cite{yu2019terminator} introduced a UI testing method, excluding the need for source code. It relies on test case descriptions and historical data for prioritization, utilizing an SVM model trained with active learning to prioritize test cases for failure detection. However, optimal performance requires multiple runs, gaining knowledge incrementally with each new executed test case.

In this paper, we model test case prioritization for UI testing as a multi-objective optimization problem, employing web page segmentation techniques. Intuitively, elements of the same level from same segments (referred to as siblings) are likely to encounter similar errors since they share the same functionality. Hence, we do not necessarily prioritize all test cases that test the function of siblings of elements from prioritized test cases. Based on that intuition, we introduce four coverage criteria focusing on objects and segments as objectives for the optimization problem and use evolutionary search algorithms to search for the optimal permutation. Different from other coverage-based approaches, our method does not rely on source code information but web page elements (buttons, links, etc.). Thus, it falls into the category of black-box TCP methods, which are more suitable in UI testing scenarios. 

We conducted experiments to evaluate our methods using two backbones (AGE-MOEA \cite{panichella2019adaptive}, NSGA-II \cite{deb2002fast}), along with other TCP methods, on a real-world dataset comprising 11 test suites. The results demonstrate that our top-performing model, AGE-MOEA, outperforms all compared methods, achieving the highest Average Percentage of Faults Detected (APFD) and APFD with Cost (APFDc) at 87.8\% and 79.2\%, respectively. Statistical tests further substantiate the significant improvements our approach offers over other methods. These findings suggest that adopting multi-objective prioritization and our new coverage criteria produces better performance in prioritizing UI test cases.

In this paper, we make the following primary contributions:
\begin{enumerate}
    \item Introduction of a multi-objective optimization approach for prioritizing UI test cases, emphasizing the diversification of segments and objects within test cases. For the first time, Web Page Segmentation is applied in the context of UI TCP.
    
    \item Construction of a new dataset of 9 subject systems with 11 test suites for the UI test case prioritization, covering a diverse range of test frameworks and frontend frameworks.
    
    \item Execution of an empirical experiment covering various aspects, wherein we compare our proposed approach to other state-of-the-art methods. The results demonstrate a significant improvement in the effectiveness of our approach.
\end{enumerate}

\section{Motivation}
\label{sec:mov-ex}
\subsection{Examples and Observations}

\begin{figure}[t]
    \centering
    \includegraphics[width=\linewidth]{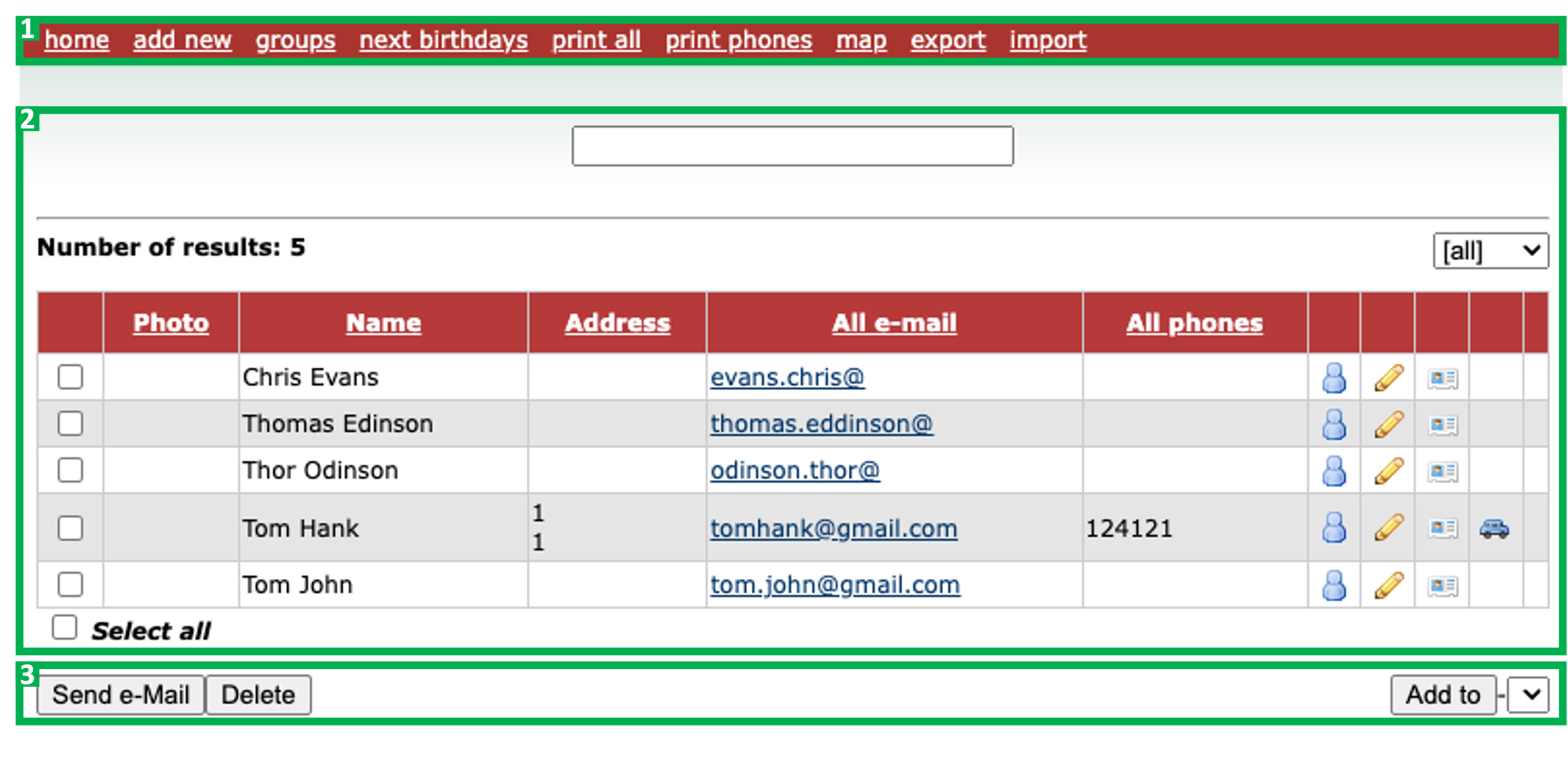}
    \caption{An example of a PHP Web application. 
    \includegraphics[height=\ht\strutbox]{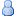}, 
    \includegraphics[height=\ht\strutbox]{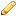}, and 
    \includegraphics[height=\ht\strutbox]{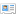} 
    denote button \textit{'details'}, \textit{'edit'}, and \textit{'print'}, respectively.}
    \label{fig:motivating_example}
\end{figure}    
Let us use a real-world Web application to illustrate the problems and motivate our approach. Figure~\ref{fig:motivating_example} shows the main page of an address book Web application written in PHP. The website allows users to edit, print, and view the details of any address record. Users can select a record and send an email to a person. It also enables them to navigate to other pages via menu~buttons.

From the example, we make the following observations.

\subsubsection*{Observation 1 [Regions in a web page]}
{\em A web page often contains different regions with respect to their functionality.}

For example, the web page from Figure~\ref{fig:motivating_example} can be divided into 3 regions according to their functional behavior and similarity (as presented by the green boxes). \textit{Region 1} presents the navigation menu of the application. \textit{Region 2} is for the information panel, providing a comprehensive display of pertinent content. Lastly, \textit{Region 3} is designated for action buttons, allowing users to execute specific tasks or commands with the selected records.

Although current UI testing prioritization methods~\cite{yu2019terminator} have achieved some success, their scope is constrained when it comes to comprehensively addressing and diversifying various sections within a web page. These methods adopt an approach that treats all objects equally when prioritizing test cases. This may result in scenarios where a test case covering numerous objects is given a higher rank, even if those objects are all part of the same region with similar functions, thereby leaving other regions untested. 
 \pagebreak
\subsubsection*{Observation 2 [Behavior of elements in the same region]}
{\em In the same region, some Web elements share the same behavior.}

\begin{lstlisting}[style=htmlStyle, caption=A snippet of HTML code for the table in Figure \ref{fig:motivating_example}., label=listing:html_addressbook]
...
<td>Chris Evans</td>
...
<td>
    <button onclick="edit(1)">Edit</button>
</td>
...
<td>Thomas Edinson</td>
...
<td>
    <button onclick="edit(2)">Edit</button>
</td>
...
<td>Thor Odinson</td>
...
<td>
    <button onclick="edit(3)">Edit</button>
</td>
...
\end{lstlisting}

Examine the HTML representation of the table in Listing \ref{listing:html_addressbook}, and observe that the recurring 'edit' buttons share identical functionality (invoking the 'edit()' function with an ID parameter) whenever a click action is triggered on them. If the 'edit' function is inadvertently modified, test cases associated with this button may fail or exhibit unexpected behavior. In such cases, fixing one function may result in the success of multiple test cases. Likewise, the 'details' and 'print' buttons are likely to follow the same behavior pattern.
Many web pages adopt this strategy, where UI components in a particular section exhibit consistent behavior. In such cases, testing a single representative element may be wise, allocating testing resources to diversify test cases for other elements with distinct behaviors.




The state-of-the-art test case prioritization approaches have not taken into account the similarity in behavior of the test objects in the same region. This could lead to the scenario that the test cases cover the objects that have been tested.

\begin{table}[t]
\centering
\caption{A sample test suite for Address Book Application. The number inside the parentheses indicates the index of the line in table (e.g. 
\includegraphics[height=\ht\strutbox]{images/icons/status_online.png} (1) is for button 'details' on the first line). For the sake of visualization, we only present the test objects visible in Figure \ref{fig:motivating_example}.}
\label{tab:sample-ts}
\resizebox{\columnwidth}{!}{%
\begin{tabular}{|l|llllll|l|l|}
\hline
\textbf{Test Case} & \multicolumn{6}{c|}{\textbf{Test objects}} & \textbf{No. Objs} & \textbf{No. Segs} \\ \hline
\textbf{TC1} & \multicolumn{1}{l|}{\scalerel*{\includegraphics{images/icons/pencil.png}}{\strut} (1)} & \multicolumn{1}{l|}{\scalerel*{\includegraphics{images/icons/pencil.png}}{\strut} (2)} & \multicolumn{1}{l|}{\scalerel*{\includegraphics{images/icons/pencil.png}}{\strut} (3)} & \multicolumn{1}{l|}{\scalerel*{\includegraphics{images/icons/vcard.png}}{\strut} (1)} & \multicolumn{1}{l|}{\scalerel*{\includegraphics{images/icons/vcard.png}}{\strut} (2)} & \scalerel*{\includegraphics{images/icons/vcard.png}}{\strut} (3) & 6 & 1 \\ \hline
\textbf{TC2} & \multicolumn{1}{l|}{\scalerel*{\includegraphics{images/icons/pencil.png}}{\strut} (1)} & \multicolumn{1}{l|}{\scalerel*{\includegraphics{images/icons/vcard.png}}{\strut} (1)} & \multicolumn{1}{l|}{\scalerel*{\includegraphics{images/icons/status_online.png}}{\strut} (1)} & \multicolumn{1}{l|}{'Select all'} & \multicolumn{1}{l|}{'Send Email'} & 'export' & 6 & 3 \\ \hline
\textbf{TC3} & \multicolumn{1}{l|}{'home'} & \multicolumn{1}{l|}{'add new'} & \multicolumn{1}{l|}{\scalerel*{\includegraphics{images/icons/status_online.png}}{\strut} (1)} & \multicolumn{1}{l|}{'Checkbox' (1)} & \multicolumn{1}{l|}{'Send Email'} &  & 5 & 3 \\ \hline
\end{tabular}%
}
\end{table}

As an illustration, consider Table \ref{tab:sample-ts}, which exhibits a test suite example for the Address Book application. In this scenario, algorithms such as Greedy Additional~\cite{li2007search} and Greedy Total~\cite{li2007search}, which exclusively rely on UI objects, assign the following ranking to the test cases: \{TC1, TC2, TC3\}. This ranking is determined by the count of test objects covered in each case. Notably, Test Case 1 (TC1) attains the highest ranking due to encompassing the most extensive array of test objects. However, it's crucial to note that the functions underlying these test objects overlap, resulting in the testing of only two distinct functions. In contrast, Test Case 2 (TC2) or Test Case 3 (TC3) may be more effective in uncovering bugs since they span a broader array of distinct regions.

\subsection{Key Ideas}

Based on the above observations, it can be seen that an effective test case prioritization method should not only take into account object coverage but also consider region coverage. This entails the ability to differentiate web objects across various regions. Additionally, the approach should be designed to circumvent scenarios where distinct objects share identical underlying functionality. We draw from the observations the following key ideas:

\
\subsubsection*{Key Idea 1: Segmentation-based Objective}
Our test case prioritization strategy considers region coverage in which the UI objects share the same or similar groups of functionality.

\subsubsection*{Key Idea 2: Multi-Objective Test Case Prioritization}
Rather than simply maximizing UI object coverage, we consider multiple objectives in test case prioritization including, diversifying test objects, covering a variety of object types (e.g., button, a tag, checkbox, input, etc.), and testing objects with similar parent structures.


\subsubsection*{Key Idea 3: Object with the same behavior}
We diversify the object coverage with different underlying functionality to
avoid redundant testing of similar functionalities.

In the Address Book example, instead of focusing solely on covering all the 'edit' buttons (which invokes the same function), we prioritize test cases that also cover the 'details' and 'print' buttons, which likely exhibit different behaviors. This multi-objective approach ensures a wider range of test scenarios while efficiently covering different objects. 

The test cases that test the rows of the table with the same underlying functionality will not be put in priority. For example, in the sampled test suite in Table~\ref{tab:sample-ts}, instead of prioritizing test cases on the multiple rows based on the number of objects covered, we also consider the diversity of regions and underlying functions. A multi-objective prioritization method would rank TC2 or TC3 higher than TC1, as they test more diverse aspects of the application and are likely to reveal more unique bugs across different regions.



\section{Segment-based Test Case Prioritization}
\begin{figure*}[t]
  \centering
  \includegraphics[width=\textwidth]{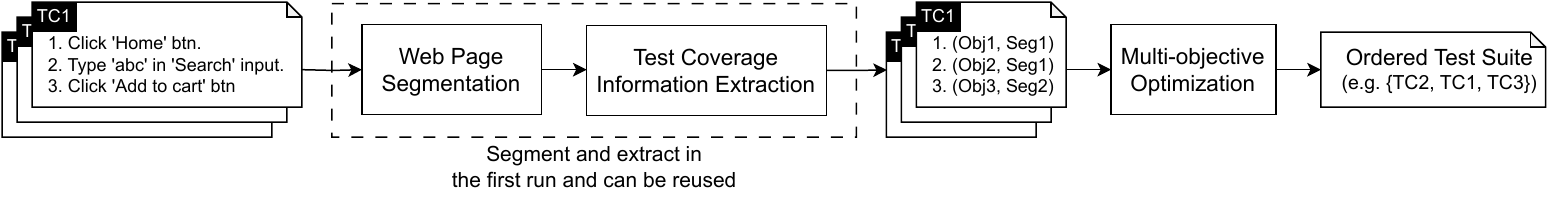}
  \caption{Overview of Segment-based Test Case Prioritization Method}
  \label{fig:method-overview}
\end{figure*}

In this section, we present our proposed method for the TCP problem using a multi-objective optimization approach.

Our primary concept involves prioritizing test cases to maximize coverage rates on four key subject objectives: segment, sibling, object type, and object. To achieve this objective, we employ a multi-objective optimization approach utilizing genetic algorithms. Similar to other TCP methods, our approach takes the test suite as input and produces an optimal execution order. This process involves two main steps: coverage information extraction and multi-objective optimization, which are visualized by Figure \ref{fig:method-overview}.

\subsection{Coverage Information Extraction}
As a coverage-based approach, we require coverage information of object and segmentation. As for object information, we need to know the list of interacting test objects for each test case. Each object information will have the information of the absolute XPath, object type (or tag name), and enclosing segment. Since the test suite alone does not provide sufficient information for this step, we need to execute the test script to get coverage data and perform web page segmentation to get segmentation data. 

\begin{definition} \textbf{(Segment)}
    A segment in a web page is a region that contains elements that have similarities in structure \cite{huynh2023web}. A segment is represented as a DOM subtree.
\end{definition}

Recapitulating Web Page Segmentation: Web page segmentation involves breaking down a web page into distinct and meaningful sections based on various factors such as layout, content, and structure. We apply a DOM-based web page segmentation approach \cite{huynh2023web} to get segmentation information for each test state. This approach takes the DOM tree of a web page as input and returns a list of DOM nodes representing the segments in that web page. This approach has demonstrated impressive results on a dataset of nearly 2000 real-world web pages.

While executing test cases, we inject the code to watch the test actions (e.g. click, type, trigger, keyboard shortcut, etc.) and changes in the HTML representation of the current app under test. Whenever an action is performed, we save the interacted HTML element with information on the URL of the current page, the object's XPath, the object's type, and other attributes. At the same time, we also perform web page segmentation to get the enclosing segment of the current test object. 

As a result, for every test case, we got the coverage information as a list of test objects associated with their containing segment. 

\subsection{Overview of Multi-objective Optimization}


\begin{figure}[t]
    \centering
    \includegraphics[width=0.65\linewidth]{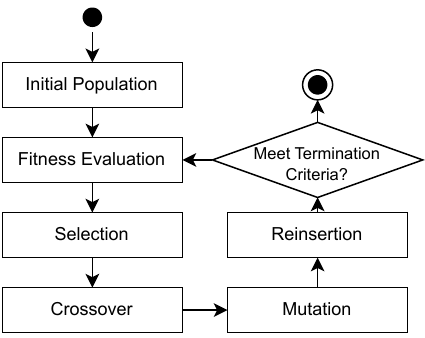}
    \caption{Overview of our genetic algorithm}
    \label{fig:genetic-algo}
\end{figure}

Genetic Algorithms (GenA) have been utilized in prior work in the scope of TCP and TCS (Test Case Selection). Figure \ref{fig:genetic-algo} shows a general process of a single-objective GenA, which includes six main steps. GenA begins by randomly generating a group of individuals, each with a unique sequence of parameters known as a chromosome. The chromosome represents a potential solution to the problem at hand. The initial population is then evaluated for their "fitness" using a specific function designed to measure how well they solve the problem. A subset of the fittest individuals is then selected to act as parents for the next generation. The next generation is created through a process of crossover, where two parents' chromosomes are combined to create offspring, and mutation, where small changes are made to the offspring's chromosome to introduce variation. This new population replaces the old one and the cycle continues until a satisfactory solution is attained or predetermined termination criteria are met.

In this work, we employ two different multi-objective genetic algorithms, known as NSGA-II and AGE-MOEA.

NSGA-II, similar to the single-genetic algorithm, follows six main steps: initialization, evaluation, selection, crossover, mutation, and reinsertion. The key difference lies in the selection phase. Since NSGA-II deals with multiple objectives, determining the fittest solution becomes challenging. To address this, it employs the principle of Pareto optimality, prioritizing solutions that are not dominated by others in the current population.

To decide which individuals to select, NSGA-II uses a crowding distance. Essentially, solutions farther from the population's center are more likely to be chosen. Additionally, the algorithm incorporates a fast non-dominated sorting algorithm to identify individuals forming the current Pareto frontier. This helps retain diverse and high-quality solutions in the next generation. Across successive generations, NSGA-II converges towards stable solutions, specifically the Pareto-optimal set for the given problem.

Pareto non-dominated solutions refer to a set of solutions in multi-objective optimization where no other solution exists that simultaneously improves all objectives; formally, a solution $\mathbf{x}^*$ is Pareto non-dominated if there is no other solution $\mathbf{x}'$ such that $f_i(\mathbf{x}') \leq f_i(\mathbf{x}^*)$ for all objectives $i$ and $f_j(\mathbf{x}') < f_j(\mathbf{x}^*)$ for at least one objective $j$, where $f_i$ represents the objective functions.

NSGA-II performs well for optimization problems with two to three objective functions \cite{deb2002fast}. However, as the number of objectives increases, the effectiveness of this algorithm tends to degrade. To resolve this problem, \cite{panichella2019adaptive} has introduced the AGE-MOEA algorithm, which builds upon NSGA-II but introduces modifications in the selection phase to better handle multi-objective optimization. 

While NSGA-II employs crowding distance for solution selection, AGE-MOEA replaces this with a survival score that accounts for both diversity and proximity, adapting iteratively to estimate the optimal Pareto front's geometry. This adaptive approach allows AGE-MOEA to maintain a balanced population and enhance the convergence towards the Pareto-optimal set, without making any assumptions about the front's shape, which results in more robust and efficient optimization outcomes.

\subsection{Multi-objective Optimization Formulation}

{\textcolor{custom-blue}{
\subsubsection{Encoding}
We expect the output solution to be an order of test cases. Given a test suite has a length of \(N\) test cases, the number of all possible solution candidates is \(N!\). In the set of candidates, let \(S\) be a random solution, \(S = \{TC_i, TC_j, \ldots, TC_k\}\), meaning that the order for execution is \(TC_i\), then \(TC_j\), and \(TC_k\) is executed last.}}

{\textcolor{custom-blue}{
\subsubsection{Crossover}
We adopt the PMX-Crossover operator, a method widely applied in previous TCP studies \cite{birchler2023single, marchetto2015multi}. This operator combines a pair of solutions by randomly selecting an intermediate point and exchanging permutation elements at that point between the two solutions.}}

{\textcolor{custom-blue}{
\subsubsection{Mutation}
Consider a chromosome $p$ with $|p| = N$. During each mutation event, the chromosome $p$ can undergo modification through one of three mutation operators: swap, insert, and invert.}}

{\textcolor{custom-blue}{
\begin{enumerate}
    \item Swap Operator:
    Randomly select two positions $i$ and $j$ from $p$ and exchange the corresponding test cases, resulting in the creation of a new offspring.
    \item Invert Operator:
    Randomly choose two positions $i$ and $j$ $(0 \leq i < j \leq N)$, then reverse the segment from $p_i$ to $p_j$ to generate a new offspring.
    \item Insert Operator:
    Randomly select two positions $i$ and $j$, and relocate $p_i$ to position $j$, generates a new offspring.
\end{enumerate}
}}

\subsubsection{Fitness functions}

{\textcolor{custom-blue}{
In our framework, the multi-objective optimization for test case prioritization is targeted toward the four types of entities: {\bf segments}, {\bf sibling objects}, {\bf object types}, and {\bf objects} as explained in Section \ref{sec:mov-ex}. First, for segments, we prioritize the ranking with the test cases covering more segments in the web pages. Second, we observe that there are objects that are different, but they share the same underlying function. Focusing on these objects exclusively may result in redundant testing. In order to prevent redundancy, we estimate repeated behaviors by grouping {\bf sibling objects} and aim to extend the coverage of sibling objects across multiple test cases, and we prioritize those test cases. }}

{\textcolor{custom-blue}{
\begin{definition} \textbf{(Sibling Objects)}
In the scope of our work, siblings are defined as objects within the same segment that share the same absolute XPath skeleton (XPath without indices).
\end{definition}}}

Third, we recognize the need to diversify more object types within the same segment, such as buttons, inputs, checkboxes, and other web elements, to improve the likelihood of detecting various faults on those elements. Lastly, we also aim to promote the coverage for more objects from the test cases in the test suite.

{\textcolor{custom-blue}{
To define the fitness function for each of the four objectives (segments, sibling objects, object types, and objects), we have drawn the inspiration from the APFD metric ~\cite{rothermel2001prioritizing} that is popularly used in evaluating the ranking of the test cases in a test suite. Rothermel {\em et al.}~\cite{rothermel2001prioritizing} introduced the APFD metric to evaluate the effectiveness of prioritized test suites {\em in fault detection}, which is widely recognized as the ultimate goal for test case prioritization in several approaches~\cite{khatibsyarbini2019test, yu2019terminator, miranda2018fast}. This metric provides insights into the capability of prioritization techniques to reveal faults sooner, i.e., by the test cases that were ranked higher. APFD is computed as follows:
\begin{equation}
    \label{apfd}
    APFD=1-\frac{TF_1+TF_2+...+TF_m}{nm}+\frac1{2n}
\end{equation}
where {\bf $TF_i$ is the first test case that reveals fault $i$}, $m$ is the total number of faults revealed by the test suite and $n$ is the total number of test cases in the test suite. APFD ranges from 0\% to 100\%, a higher value means better ordering of test cases in terms of early fault detection.  That is, {\em APFD gives a higher score for the ranking of the test cases in a test suite in which more test cases {\bf revealing the faults} are ranked higher}.
}}

{\textcolor{custom-blue}{
Inspired by the spirit of APFD, we introduce the fitness function for each objective as follows. Let us take the segment objective as an example. The fitness function for the segment objective, $\mathcal{F}_{seg}$, gives more rewards to a ranking of test cases in a test suite such that more test cases {\bf covering the segments} in a web page are ranked higher. Mathematically,    
\begin{equation}
\label{fitness-seg}
\centering
\mathcal{F}_{seg} = 1 - \frac{TS_1 + TS_2 + \ldots + TS_m}{nm} + \frac{1}{2n}
\end{equation}
where $m$ denotes the number of segments, $TS_i$ denotes the {\em first test case in the test suite that covers the segment $i$}. The formula (\ref{fitness-seg}) means that we aim to cover our more segments in the ranked list of test cases, expecting that the test suite can detect faults sooner.}}

{\textcolor{custom-blue}{
Note that, the APFD metric (Formula (\ref{apfd})) and the fitness function (Formula (\ref{fitness-seg})) differ from each other in the following aspects. First, while both APFD and $\mathcal{F}_{seg}$ are the measurements of the quality of a ranking of test cases in a test suite, APFD favors the ranking with {\em more high-ranked test cases revealing faults} and $\mathcal{F}_{seg}$ favors the one with {\em more high-ranked test cases covering the segments}. The intuition is that covering more segments early in the ranked list of test cases would lead to revealing faults earlier. Second, APFD requires the knowledge of the ground truth on the faults, while $\mathcal{F}_{seg}$ does not because it focuses on the coverages of the web segments. Third, solely covering more segments does not always lead to revealing the faults earlier because in our framework, we aim to have multi-objective optimization with objectives other than covering segments. In brief, {\em when we use {\bf APFD as the evaluation metric for our framework (using revealing faults as a criterion for evaluation}), the evaluation is independent of the {\bf fitness functions (using coverage on segments as a criterion for a fitness function) in our approach}}. 
}}

{\textcolor{custom-blue}{
Similar to Formula (\ref{fitness-seg}), we define fitness functions for the other three objectives: 1) $\mathcal{F}_{sib-objects}$ is the average percentage sibling object coverage , 2) $\mathcal{F}_{obj-types}$ is the average percentage object type coverage, and 3) $\mathcal{F}_{obj}$ is the average percentage object coverage.}}

{\textcolor{black}{
Since the object type is the basis forming other types of entities, a test suite that cover all objects inherently guarantees coverage of the other objective elements (segments, sibling objects, and object types). However, optimizing the coverage of objects $\mathcal{F}_{obj}$ does not guarantee that other functions ($\mathcal{F}_{seg}$, $\mathcal{F}_{sib-objects}$, $\mathcal{F}_{obj-types}$) will also be optimized.
There is an interdependent relationship among these functions: increasing the coverage of segments $\mathcal{F}_{seg}$, sibling objects $\mathcal{F}_{sib-objects}$, or object types $\mathcal{F}_{obj-types}$ naturally increases the coverage of objects $\mathcal{F}_{obj}$. On the other hand, $\mathcal{F}_{sib-objects}$ and $\mathcal{F}_{obj-types}$ do not have such direct proportionality. $\mathcal{F}_{sib-objects}$ and $\mathcal{F}_{obj-types}$ both depend on segments and objects but differ in their nature. Maximizing one function does not lead to the maximization of the other function. Therefore, it is not possible to have a single solution that can optimize all four objective functions simultaneously.
As a result, the responsibility of the Genetic Algorithm is to find a test case order that optimizes the trade-off between these fitness functions.
}}



{\textcolor{custom-blue}{
\subsubsection{Choosing a Pareto optimal solution}
Multi-objective optimization algorithms return a set of non-dominated solutions on the Pareto front, revealing trade-offs between their respective objective functions. Each candidate solution has four values for four aforementioned fitness functions. The next step in our method involves determining the most optimal solution among the set of non-dominated solutions. We employ multi-criteria sorting on the set of non-dominated solutions, with the sorting weights in the following objective order: $\mathcal{F}_{seg}$ (segments), $\mathcal{F}_{sib-objects}$ (sibling objects), $\mathcal{F}_{obj-types}$ (object types), and $\mathcal{F}_{obj}$ (objects). 
}}


\section{Empirical Evaluation}

\subsection{Research Questions}
\begin{enumerate}[label=\textbf{RQ\arabic*}]
    \item \textbf{(Fault Detection)} How does our method compare with other existing TCP approaches in terms of average fault detection rate (APFD)?
    {\textcolor{custom-blue}{
    \item \textbf{(Fault Coverage)} How many test cases are needed to fully cover all faults in a project in the oracle?}}
    \item \textbf{(Redundant Testing)} How does our method compare with other existing TCP approaches in terms of diversifying test objects to prevent redundant testing?
    {\textcolor{custom-blue}{
    \item \textbf{(Time Efficiency)} How time efficient is our approach in comparison with the baseline?
    }}
\end{enumerate}

\subsection{Baseline Approaches}

\textbf{Random-based approaches:}

\textit{Random order:}  In this approach, test cases are ordered randomly. Random Prioritization is the most straightforward in concept and the simplest and most cost-effective in implementation \cite{zhou2020beating}. As the most basic method, it serves as the lower bound in our experiment. We apply this technique to each test suite $n$ times, with $n$ being the number of test cases of test suite, and record the evaluation values each time the ordered test suite is executed. These values are then averaged to obtain the final results.

\textit{Zhou et al. \cite{zhou2012fault} [\textbf{ART-F}]:} This technique is a family of ART-based TCP methods guided by coverage and similarity. ART \cite{jiang2009adaptive, chen2010adaptive}, designed to improve random testing, operates based on the principle of evenly spreading test cases across the input domain. ART-F is one of the implementations based on this principle. At each iteration, a fixed-size "candidate set" is dynamically formed by randomly selecting a test case from the set of not yet prioritized test cases. The test case with the maximum Manhattan distance from the already executed test cases is chosen next in the prioritized suite.

\noindent \textbf{Greedy approaches:} 

\textit{Greedy Total \cite{rothermel2001prioritizing} [\textbf{GT}] :} This approach orders test cases with respect to the number of entities covered by each test case. The total number of entities covered by each test case is counted and the test suite is then ordered in descending order of that number.

\textit{Greedy Additional \cite{rothermel2001prioritizing} [\textbf{GA}] :} This approach is similar to GT, the sole difference being that it selects the next test case as the one that covers the most entities among those not yet covered by the prioritized test cases.

\textit{Additional Spanning \cite{miranda2018fast} [\textbf{GA-S}] :} At each iteration, it selects the test case that covers most entities not yet covered among those in "spanning set". The concept of "spanning set" for coverage criterion is first introduced in \cite{marre2003using}. A spanning set is a minimum subset of entities with the property that any set of test cases covering this subset covers every entity in the program.

\noindent \textbf{Machine Learning-based approaches:} 

\textit{Yu et al. \cite{yu2019terminator} [\textbf{Terminator}]:} 
Terminator is a test case prioritization method specifically designed for UI testing. It employs a Support Vector Machine (SVM) model, using input from any of text descriptions of test cases, results from previous runs (i.e. test cases failed, passed or skipped), or a combination of both. However, in our experiment, we only use text description as the sole input due to the lack of historical data, which is introduced as Terminator-F1. The SVM model is trained through active learning and distinguishes between passed and failed test cases using certainty sampling and uncertainty sampling strategies. Notably, the method is deemed real-time as it continuously acquires knowledge with each execution of a test case.

\subsection{Evaluation Metrics} 

\subsubsection{APFD}

The first metric is the Average Percentage of Faults Detected (APFD)~\cite{rothermel2001prioritizing}. APFD ranges from 0\% to 100\%, a higher value indicates better ranking of test cases in early fault detection.


\subsubsection{APFDc}\label{sec:apfdc}
APFD does not take into account the execution time and the severity of test cases while in reality, some test cases might take more time to execute compared to others, and some faults might cause more severe damage if encountered. As such, Elbaum {\em et al.}~\cite{elbaum2001incorporating} propose APFDc, which treats test cases differently based on the cost and severity of test cases. APFDc is calculated as follows:
\begin{equation}
    \label{apfdc}
    APFDc=\frac{\sum_{i=1}^{m}(f_i\times(\sum_{i=TF_i}^{n}t_i-0.5t_{TF_i}))}{\sum_{i=1}^{n}t_i\times\sum_{i=1}^{m}f_i}
\end{equation}
where $t_1, t_2, ... t_n$ are the cost for $n$ test cases, $f_1, f_2, ... f_m$ are the severity of $m$ faults and $TF_i$ is the first test case that reveals fault $i$. In this work, we only take into account the cost of test cases and ignore the severity of faults (i.e. $f_1 = f_2 = ... = f_m = 1$). 

\subsubsection{NAPFD}\label{sec:napfd}
APFD and APFDc both assume that ranked test cases can detect all faults. Yet, the cost for running the whole test suite is expensive while the budget is limited. In such situations, there is a likelihood of variations in the test suite of each run, leading to the failure in detecting all faults~\cite{pan2022test}. Thus, Qu {\em et al.} \cite{qu2007combinatorial} proposed NAPFD, which can be used in scenarios where the entire test suite is not executed. The formula is as follows:
\begin{equation}
    \label{napfd}
    NAPFD=p-\frac{TF_1+TF_2+\cdots+TF_{m}}{nm}+\frac{p}{2n}
\end{equation}
where $p$ is the number of faults detected by the prioritized test suite divided by the number of faults detected in the full test suite. When $p =  1$, the NAPFD becomes APFD. 

\subsubsection{MTFD}\label{sec:mtfd}
Aside from the aforementioned metrics, we also measure the average percentages of test execution needed to achieve 100\% of fault coverage as in \cite{nguyen2021rltcp} by computing Minimal Tests for Fault Detection (MTFD) as follows: 
\begin{equation}
    \label{mtfd}
    \centering
    MTFD = \frac{\max\{TF_1, TF_2, ... TF_m\}}{m}
\end{equation}

\subsubsection{FDR}
\label{sec:fdr}

In addition to evaluating faults detection rate, we aim to assess the effectiveness of our method in minimizing test redundancy. Test redundancy in test case selection refers to repeatedly testing of certain functions. However, in the context of prioritization, we need to execute all test cases and not ignore any of them. To assess the test redundancy in the current context, we propose a new metric called Function Duplication Rate or FDR.
\begin{equation}
\displaystyle FDR\ =\ \frac{\sum _{i=0}^{k} |F_{i} |-|\bigcup _{i=0}^{k} F_{i} |}{\sum _{i=0}^{n} |F_{i} |-|\bigcup _{i=0}^{n} F_{i} |}
\end{equation}
where \( n \) represents the total number of test cases, \( k \) is the number of test cases needed to cover all testing functions, and \( F_i \) denotes the set of functions covered by the \( i \)-th test case. A lower FDR indicates fewer duplicated functions being tested. If a solution's \( k \) value is low, indicating that fewer test cases are needed to cover all functions of test objects, it suggests a more efficient solution in minimizing test redundancy. When we need to execute all test cases to cover all functions, then \( k \) is \( n \) and \( FDR \) reaches 100\%.

\subsection{Dataset}
\label{sec:dataset}

\begin{table}[t]
\caption{Summary of test suites including the number of test cases, faulty test cases, root causes, and collected states.}
\label{table:dataset}
\begin{tabular}{c|cccc}
\textbf{Test suite}      & \textbf{TCs} & \textbf{F. TCs} & \textbf{RCs} & \textbf{States} \\ \hline
\textbf{addressbook}     & 93           & 10              & 2                  & 386             \\
\textbf{claroline}       & 235          & 8               & 1                  & 1489            \\
\textbf{dimeshift}       & 174          & 140             & 4                  & 832             \\
\textbf{pagekit}         & 37           & 10              & 2                  & 213             \\
\textbf{phoenix}         & 314          & 11              & 5                  & 1618            \\
\textbf{ppma}            & 62           & 37              & 3                  & 308             \\ \hline
\textbf{juice\_shop\textsubscript{(1)}} & 92           & 11              & 11                 & 284             \\
\textbf{juice\_shop\textsubscript{(2)}} & 92           & 57              & 4                  & 284             \\
\textbf{mattermost\textsubscript{(1)}}  & 221          & 27              & 9                  & 1029            \\
\textbf{mattermost\textsubscript{(2)}}  & 191          & 9               & 9                  & 869             \\ \hline
\textbf{moodle}          & 27           & 14              & 6                  & 413            
\end{tabular}%
\end{table}

Table \ref{table:dataset} provides a summary of 11 test suites utilized in our experiments. These test suites vary in the number of test cases, test states, root causes, and failure rates,  thereby presenting a diverse and comprehensive set of scenarios. The data comes from two sources:

\subsubsection*{Yandrapally et al. \cite{yandrapally2022fragment}} This research introduces a model-based test generation technique called Fraggen, capable of generating oracles for test cases it generates. They also provide a dataset for their experiments, including test suites for eight widely-used subject systems. To reuse these test suites, we conducted two additional steps for data preprocessing. Fraggen identifies whether a test case passes or fails by comparing the visual and DOM aspects of each state, without detailing the specific error. Therefore, we first manually categorized errors into root cause groups by reviewing captured screenshots. Subsequently, we performed web page segmentation on the recorded HTML files to derive coverage information for each test case. We selected six out of the eight projects that provided sufficient information: 'addressbook,' 'claroline,' 'dimeshift,' 'pagekit,' 'phoenix,' and 'ppma.' These subject systems span a diverse array of frontend frameworks such as React, Vue, Angular, etc.

\subsubsection*{Self-collected datasets}
We selected two open-source web applications with UI test cases: 'mattermost'~\cite{mattermost} and 'juice-shop'~\cite{juice-shop}. Mattermost is a team communication management system, while juice-shop is a web application for security training. Both projects provide end-to-end (e2e) UI test cases written in Cypress.
For Mattermost, with its multiple frontend versions, we aimed to simulate software evolution. We selected a frontend version and executed the e2e test case of the preceding minor version. Two pairs of versions were chosen: Test case v7.7.4, front-end v7.8.12 (\textbf{mattermost\textsubscript{(1)}}), and Test case v7.8.12, front-end v7.9.3 (\textbf{mattermost\textsubscript{(2)}}). Concerning JuiceShop, which released multiple versions with primary changes in the backend, we decided to replicate modifications in evolutionary applications through mutation analysis. We conducted an analysis of the e2e test cases, randomly selecting two sets of objects that have lengths of 4 and 10, causing 10-40\% of tests to fail. Mutations were performed by modifying the IDs of these test objects, resulting in two versions of the frontend application. Subsequently, the e2e test suite was consecutively executed on each frontend version to obtain two test run results.

Additionally, we adopted a test suite from \cite{nguyen2021rltcp} written in Katalon Groovy for the Moodle project, a course management system. This inclusion broadens our dataset to a total of 11 test suites.

\subsection{Experimental Setup}
\subsubsection{Experimental Environment}
We carried out the experiment on the macOS environment equipped with a a 2 GHz Quad-Core Intel Core i5 processor and 16GB of RAM. 

\subsubsection{Configuration for our algorithms} 
We tried different configurations for two backbones and only recorded results of the best configuration as the final experimental results. After exploring various configurations, we determined that the optimal settings for two evolutionary search algorithms, including AGE-MOEA and NSGA-II, were as follows: population size set to 100, the number of generations at 200, and a crossover probability of 0.5. For simplicity in the experiment results section, we use the labels \textbf{SegTCP} and \textbf{SegTCP*} to refer to our multi-objective method with AGE-MOEA and NSGA-II, respectively.

\subsubsection{Configuration for Terminator}
In the absence of historical data, we choose the F1 version of Terminator, utilizing text features as input. Terminator involves two parameters: the batch size \( N_1 \) and the threshold of the query strategy \( N_2 \). Following its paper, we set \( N_1 \)=10 and \( N_2 \)=30. For text feature extraction, we employed TF-IDF, a widely used method in many other applications. 
\section{Experimental Results}

\subsection{Comparison with Other TCP Approaches on Fault Detection Rate (RQ1)}
\label{RQ1}
\begin{table}[h]
\begin{threeparttable}
    
\centering
\caption{Average APFD, APFDc, MTFD, FDR of our methods (SegTCP and SegTCP*) and the baselines (RQ1 and RQ2)}
\label{tab:apfd-table}
\begin{tabular}{|c|c|c|c|c|}
\hline
\textbf{Approach} & \multicolumn{1}{c|}{\textbf{APFD}} & \multicolumn{1}{c|}{\textbf{APFDc}} & \multicolumn{1}{c|}{\textbf{MTFD}} & \multicolumn{1}{c|}{\textbf{FDR}} \\ \hlineB{3}
\textbf{SegTCP} & \textbf{87.8/11.5} & \textbf{79.2/17.7} & \textbf{22.5/23.3} & \textbf{45.2/32.1} \\ \hline
\textbf{SegTCP*} & 84.6/13.5 & 75.9/19.8 & 31.1/29.1 & 48.9/35.3 \\ \hlineB{3}
\textbf{GA} & 83/14.1 & 71.9/18.9 & 33.7/26.2 & 70.5/27.5 \\ \hline
\textbf{GA\_S} & 77.9/15.8 & 73.9/19.1 & 42.7/28.5 & 70.1/29.6 \\ \hline
\textbf{GT} & 76.9/12.8 & 62.4/17.6 & 46.3/27.4 & 97.3/3.1 \\ \hline
\textbf{Terminator} & 76.5/16.2 & 73.1/20.1 & 42.6/31.7 & 84.3/21.3 \\ \hline
\textbf{ART-F} & 74.1/15.5 & 74.5/15.6 & 53.1/32.3 & 94.7/7.2 \\ \hline
\textbf{Random} & 73.9/15.5 & 73.7/15.5 & 53.1/33.7 & 95.5/6.3 \\ \hline
\end{tabular}%

\begin{tablenotes}
    \small
    \item Results are presented in the format of ${M}/{\sigma}$, where ${M}$ denotes the mean value of the respective metric, and ${\sigma}$ represents the standard deviation.
    
    \item The bold values show the optimal results for the respective metrics, with the highest values for APFD and APFDc and the lowest value for MTFD (Minimal Tests for Faults Detection) and FDR (Function Duplication Rate)
\end{tablenotes}

\end{threeparttable}

\end{table}

\begin{figure}
    \centering
    \includegraphics[width=1\linewidth]{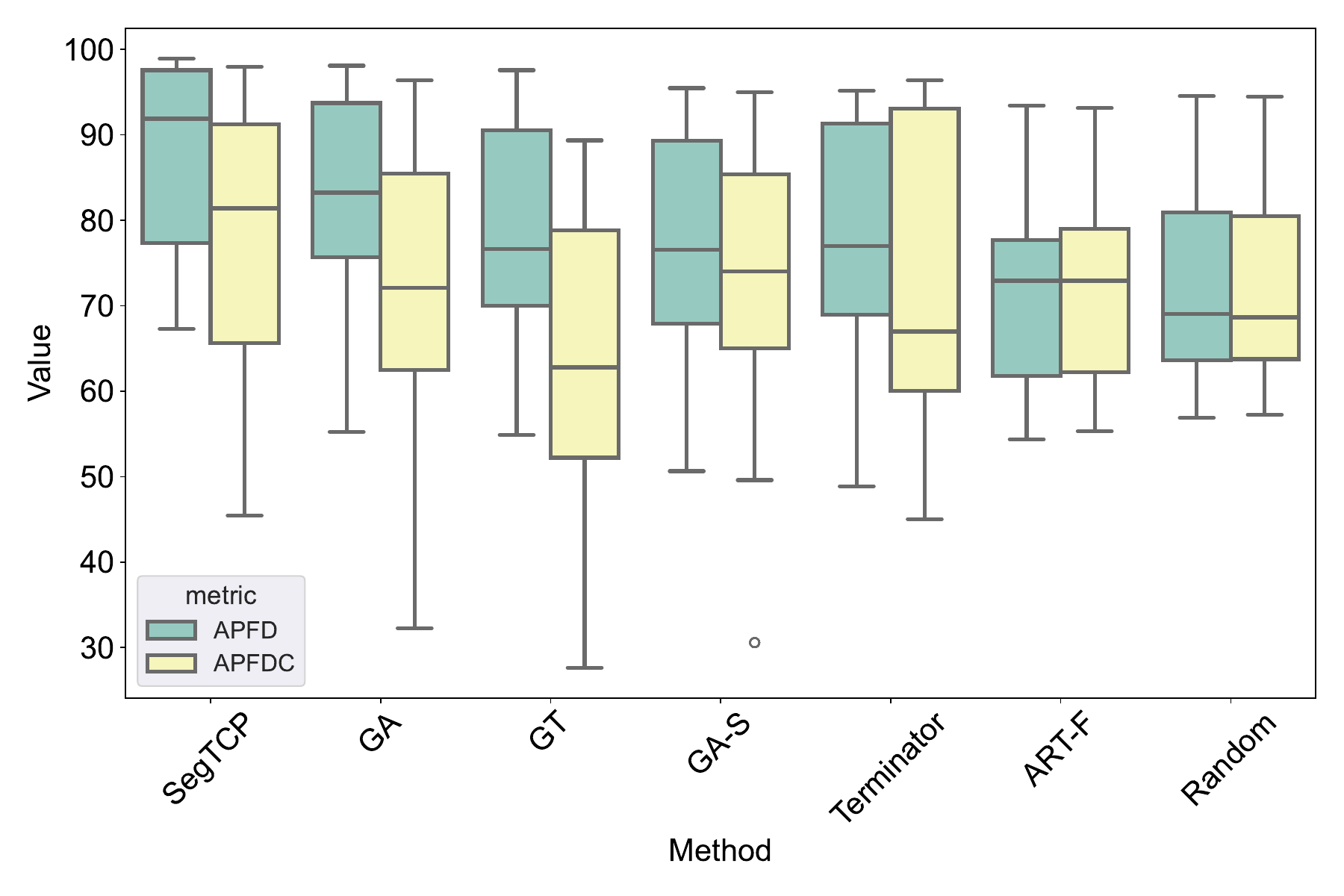}
    \caption{The APFD and APFDc distribution of our SegTCP method and other six methods over 11 test suites (RQ1).}
    \label{fig:boxplot}
\end{figure}

\begin{figure*}[t] 
  \centering
  \subfloat[Fault coverage.]{\includegraphics[width=0.33\textwidth]{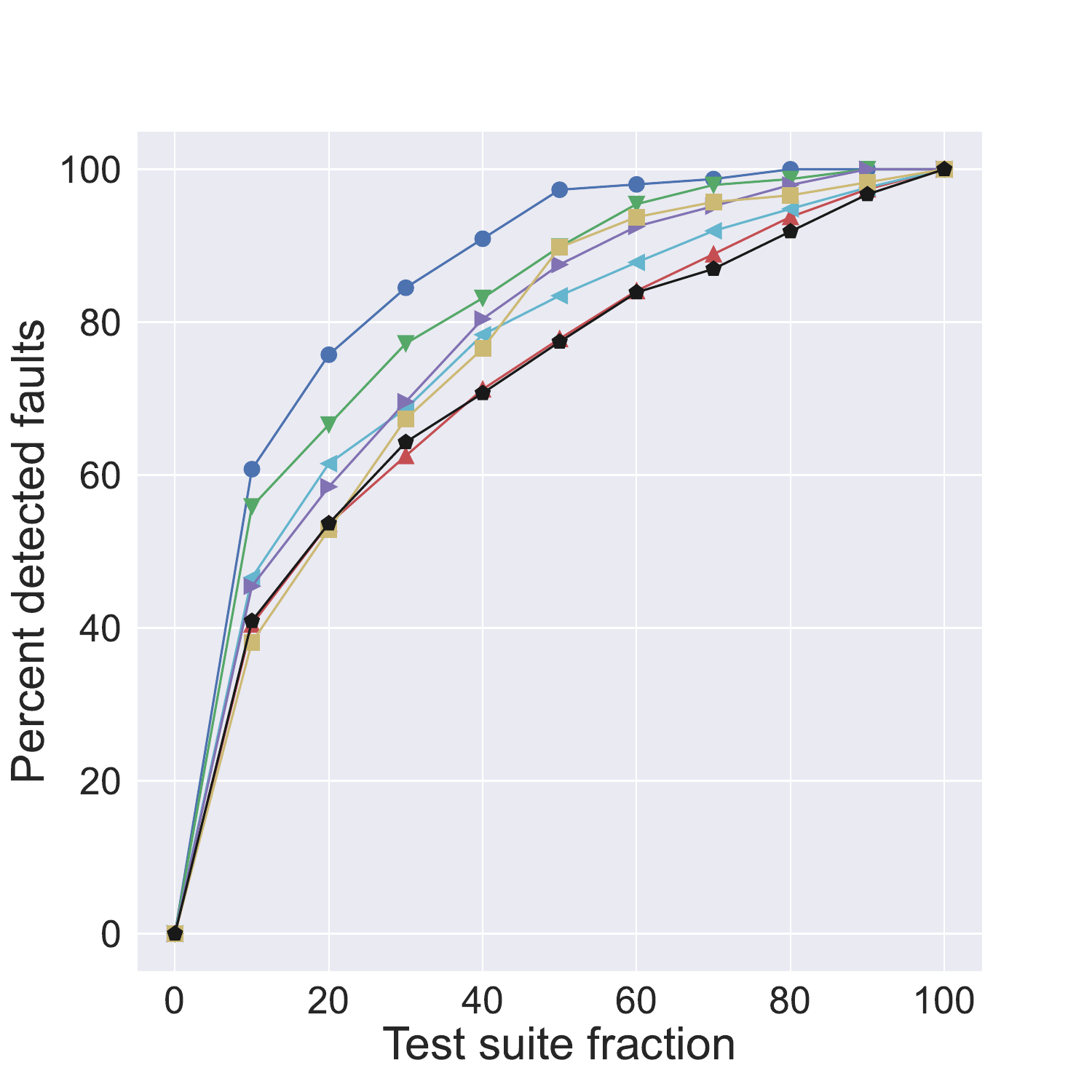}\label{fig:a}}\hfill
  \subfloat[Segment coverage.]{\includegraphics[width=0.33\textwidth]{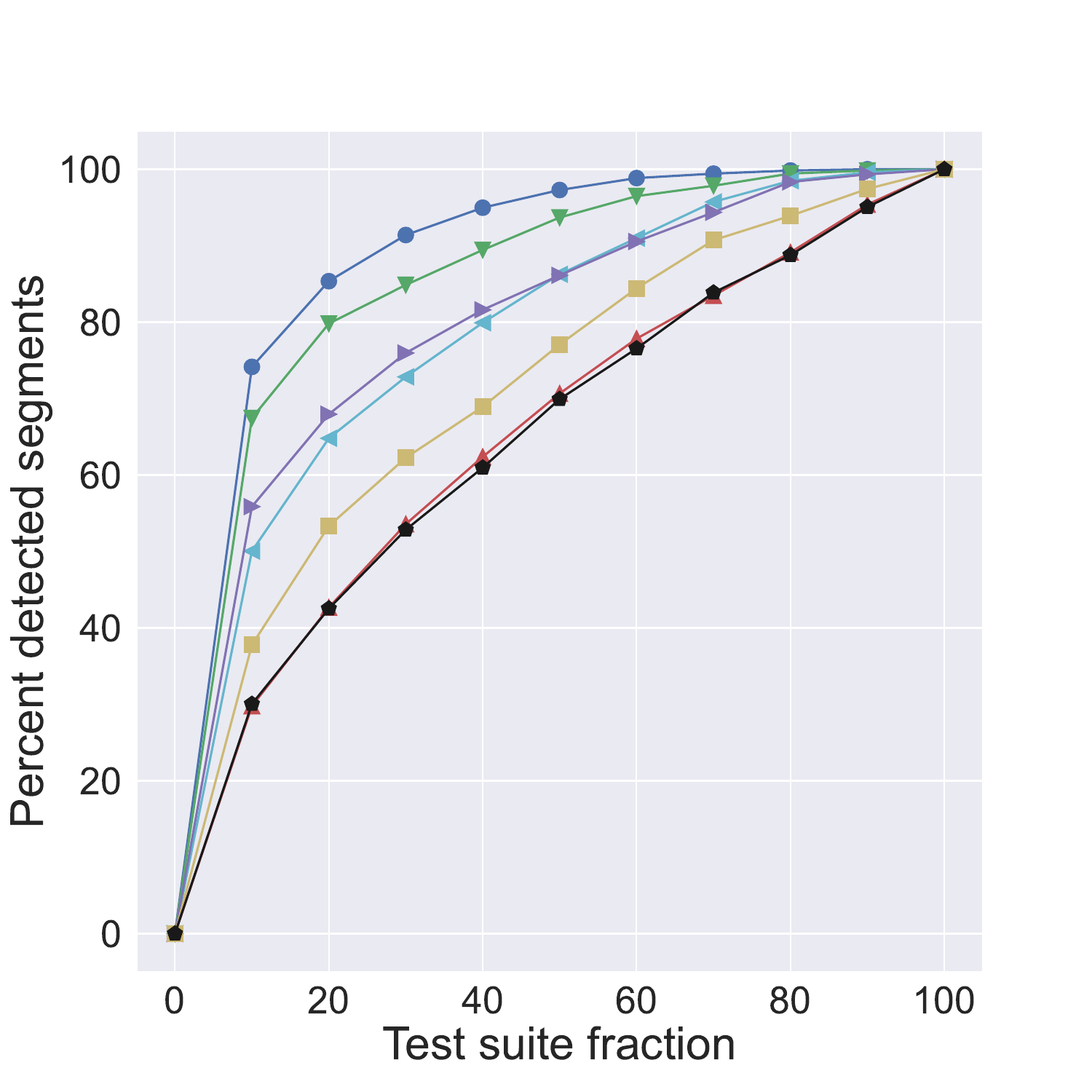}\label{fig:b}}\hfill
  \subfloat[Object coverage.]{\includegraphics[width=0.33\textwidth]{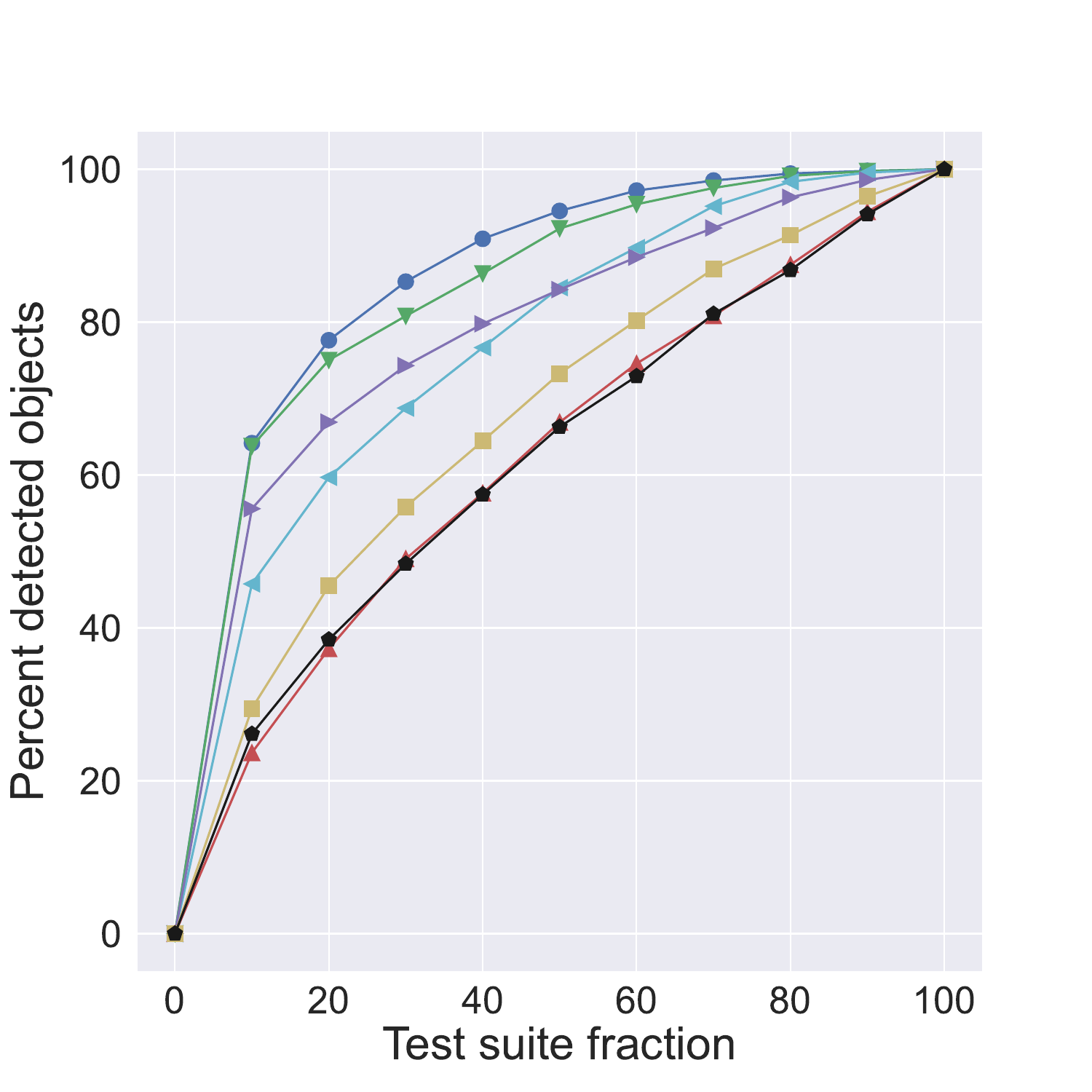}\label{fig:c}}
  
  \subfloat[Object-type coverage.]{\includegraphics[width=0.33\textwidth]{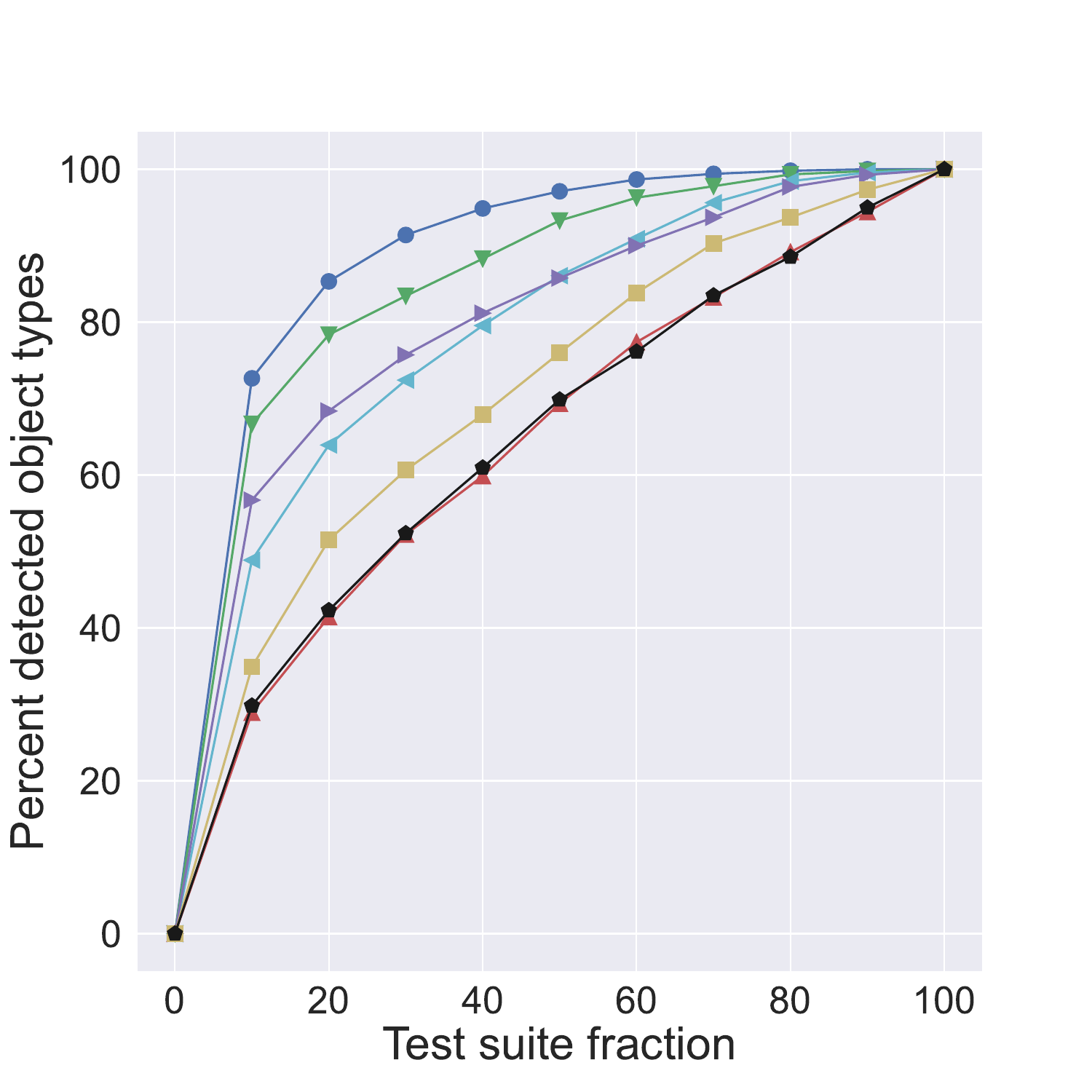}\label{fig:d}}\hfill
  \subfloat[Sibling coverage.]{\includegraphics[width=0.33\textwidth]{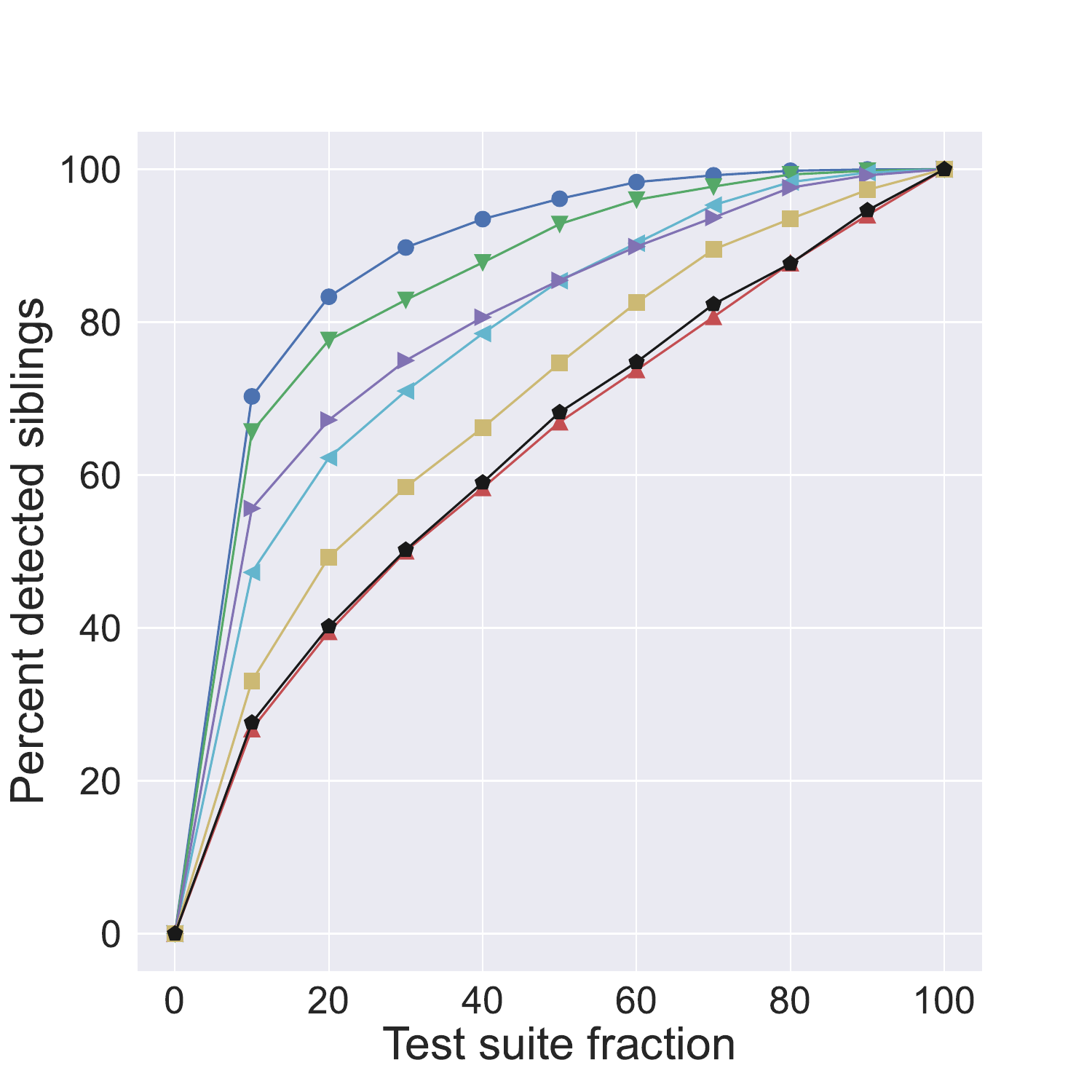}\label{fig:e}}\hfill
  \subfloat[Average NAPFD.]{\includegraphics[width=0.33\textwidth]{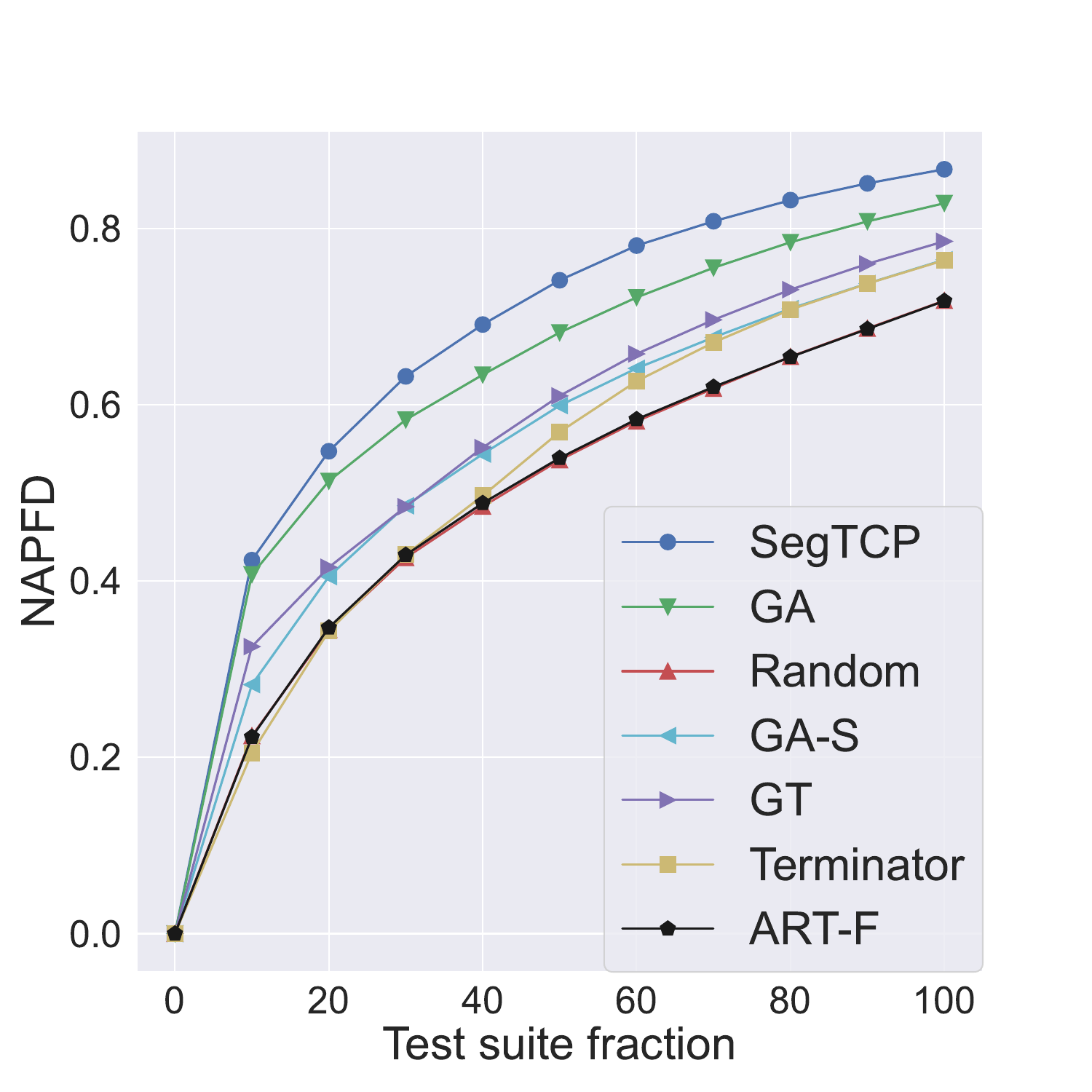}\label{fig:f}}

  \caption{Average coverage graphs for 5 types of subjects along with the Average NAPFD chart (\hyperref[RQ1]{RQ1}, \hyperref[RQ2]{RQ2})}
  \label{fig:all}
\end{figure*}

In addressing the RQ on fault detection effectiveness, we examine the APFD and APFD considering the cost (APFDc) (more details about these metrics are in Section \ref{sec:apfdc}) metrics for our proposed methods (SegTCP and SegTCP*) against six existing TCP approaches over 11 test suites as mentioned in Section \ref{sec:dataset}. The results, as outlined in Table \ref{tab:apfd-table} and visualized in Figure \ref{fig:boxplot} and Figure \ref{fig:f}, indicate that our methods demonstrate superior performance, particularly the SegTCP method, which shows the highest fault detection rates in both APFD (87.8\%) and APFDc (79.2\%) metrics. 

When it comes to greedy-based methods (GA, GA-S, GT), while they are straightforward and have a long history of usage, they still yield satisfactory results in UI TCP. Terminator-F1, on the other hand, is an ML-based approach dedicated for TCP in UI testing, demonstrates less favorable performance compared to the greedy family. The remaining two approaches, characterized as purely random and adaptive random (ART-F), produce marginal results, with the latter showing a slightly better performance.

Looking closely at the APFD results, we observe that SegTCP outperforms other approaches, with scores consistently higher than all the competing methods. This suggests that SegTCP can detect faults more quickly than the alternative TCP strategies. SegTCP* also performs well with an APFD score of 84.6\%, which is still above the majority of the competing methods, further supporting the efficacy of our approach.

The Greedy Additional (GA) approach has consistently proven itself as a strong baseline in prior research efforts. FAST \cite{miranda2018fast} and AGA \cite{li2021aga} are proposed to improve the time efficiency of GA without showing a significant difference in effectiveness. Despite the simplicity of concept and long-standing usage of Greedy algorithms, they continue to yield satisfactory results in test case prioritization. The core idea of GA is that longer test cases have a higher potential of detecting failures. This idea has been proven by the impressive APFD value of 83\%, which beats all other alternative methods except ours. However, longer test cases come with more cost of execution, contradicting the principle of APFDc. Consequently, their APFDc  are even marginally lower than those by the Random method.


The experiment considers the normalized APFD, termed NAPFD (details in Section \ref{sec:napfd}), to assess the effectiveness of TCP approaches within constrained execution time. Average NAPFDs for each TCP technique are depicted in Figure \ref{fig:f}. For a clear view, we represent our methods by the best of them which is SegTCP. 

In the initial phase, comprising less than 10\% of the test suite, our NAPFD closely aligns with that of GA. However, as the number of executed tests increases, NAPFD demonstrates a progressively prominent performance. The key reason is that at the early stage, both our method and GA tend to choose test cases with a high number of steps, which can cover many objects, segments, siblings, etc. In the latter test case, when there are more test objects, the diversity strategy of those algorithms becomes clearer and has a major difference from each other. 

Turning to Terminator, its NAPFD for test suite fractions below 40\% closely resembles random-based methods. In the early stages, Terminator is required to execute a number of random tests and construct a substantial knowledge base. This knowledge is used to train an SVM model to predict which test cases may fail and should be added to the execution queue. Terminator's enhanced performance becomes more apparent in later test executions, exceeding 50\% of the test suite.

Considering the statistical difference in APFD values between methods, Figure \ref{fig:boxplot} illustrates the APFD and APFDc values for each method across 11 test suites. SegTCP exhibits a comparatively higher median in APFD than most of the other methods, indicating that it tends to detect faults more efficiently on average. The distribution of APFDc also suggests that when considering the cost, SegTCP remains effective.



{\textcolor{custom-blue}{Figures \ref{fig:b} to \ref{fig:e} illustrate how our multi-objective goal is achieved. Our primary objective is to promptly cover all segments, siblings, object types, and objects in the test execution sequence. This achievement is depicted in the line chart. For example, Figure~\ref{fig:b} displays the percentage of the total number of segments that are covered over the gradually increasing fraction of a test suite.
As seen, only approximately 20\% of test cases are required to cover 80\% of all subject elements, and 60\% to cover 100\% of them. Importantly, referring to chart \ref{fig:a}, there is a direct impact between the element coverage and fault detection, showcasing the efficacy of our multi-objective element coverage approach in fault detection.}}

\subsection{Comparison Results with other TCP Approaches on Fault Coverage (RQ2)}
{\textcolor{custom-blue}{
The MTFD metric (see Section \ref{sec:mtfd}) quantifies the minimum proportion of the test suite necessary to cover 100\% of faults. The column MTFD in Table~\ref{tab:apfd-table} shows the results for all the approaches under study.
As seen, across 11 test suites for all 11 projects in our dataset, our approach SegTCP demonstrates a mean MTFD of 22.5\%, indicating that SegTCP requires only 22.5\% of the test cases to be executed to uncover all faults, with a standard deviation of 23\%. The approach requires testing a smaller portion of the test suite to uncover all faults than do the other approaches, compared with 33.7\% and 53.1\% by the GA and random methods, respectively.}}

{\textcolor{custom-blue}{
GA exhibits a mean requirement of one-third of the test suite to uncover all faults, which is also noteworthy. In the case of the Random method, on average, approximately a half of test suites need to be executed to cover all faults. Because when test case order is randomized, failure occurrences may distribute across various positions. In our experiment involving 11 test suites, we conducted multiple iterations for Random (n times, where n represents the number of test cases in each suite) and get the mean value of MTFD. As a result, the average percentage of test cases required to identify all failures approaches 53\%. The adaptive random method ART-F yields results similar to pure randomness.}}

{\textcolor{custom-blue}{
 Terminator, however, achieves an MTFD value of 42.6\% with a standard deviation of 31.7\%, indicating inconsistency across different test suites. Terminator's effectiveness relies on executing a sufficient number of test cases to train a robust SVM model. If the initially chosen random test cases lack a balance representation of both successful and failed test cases, it can introduce bias into the training data, potentially impacting performance. Despite outperforming pure random methods in terms of MTFD, Terminator's efficacy is still contingent upon randomness.}}

\label{RQ2}

\subsection{Comparison of TCP Approaches on Test Redundancy (RQ3)}
\label{RQ3}

Our objective here is to demonstrate the effectiveness of our method in preventing test redundancy, using the FDR from Section \ref{sec:fdr}. FDR indicates the function duplication rate of a prioritized test suite. A lower FDR points out a less redundant order of test cases.

Our methods show good results with the best FDR of SegTCP is 45.2\%, significantly outperforming greedy, ML and random-based methods. One of our objectives is maximizing the sibling's coverage rate. Given that siblings are objects in the same segment with the same xpath's skeletons. Our observation on many web pages shows that sibling elements tend to share their behavior. Diversifying siblings consequently reduces the redundancy in test functions. 

As for Greedy Additional, its ultimate strategy is to cover all objects as soon as possible, which inadvertently prevents the repeated test function, as shown by the FDR value of 70.5\%. GA-S, a variant of Greedy Additional, slightly outperforms its predecessor by incorporating the concept of a spanning set, specifically designed to reduce repeated testing elements.

The Terminator method, which is not a coverage-based TCP technique, generates prioritized test suites with a higher duplication count, as reflected in its FDR of 84.3\%. This outcome was expected, as the Terminator uses an SVM to predict and pick the test case with the highest probability of failing. This approach leads to the selection of test cases that are close or similar to previous failed test cases, thereby escalating the number of repeated testing functions.

In random-based approaches, the absence of a rule for test cases orders means that random methods continue running until the end of the execution queue to cover all testing functions. Therefore, the duplication rates for these approaches tend to approach 100\%.

{\textcolor{custom-blue}{
\subsection{Time Efficiency (RQ4)}}
\label{sec:time}

\begin{table}[t]
\centering
\small
\tabcolsep 3pt
\caption{{\textcolor{custom-blue}{Time Efficiency Comparison (RQ4)}}}
\label{tab:time}
\begin{tabular}{|c|c|c|c|}
\hline
Method     & \begin{tabular}[c]{@{}c@{}}Avg. Prioritiz.\\ Time (s)\end{tabular} & \begin{tabular}[c]{@{}c@{}}Avg. Time to Detect\\ All Faults (s)\end{tabular} & \begin{tabular}[c]{@{}c@{}}Avg. Time to Detect \\ One Fault (s)\end{tabular} \\ \hlineB{3}
SegTCP     & 75.40                                                          & 935.24                                                                         & 188.12                                                                          \\ \hline
SegTCP*    & 72.82                                                          & 1,195.17                                                                       & 220.91                                                                          \\ \hlineB{3}
GA         & 0.01                                                           & 1,309.04                                                                       & 231.35                                                                          \\ \hline
GA-S       & 0.07                                                           & 1,202.16                                                                       & 243.15                                                                          \\ \hline
GT         & 0.01                                                           & 1,799.10                                                                       & 463.15                                                                          \\ \hline
Terminator & 0.12                                                           & 1,273.25                                                                       & 223.25                                                                          \\ \hline
ART-F      & 0.01                                                           & 1,276.09                                                                       & 263.83                                                                          \\ \hline
Random     & 0.01                                                           & 1,367.93                                                                       & 268.96                                                                          \\ \hline
\end{tabular}
\end{table}

{\textcolor{custom-blue}{
\subsubsection{Prioritization Time and Fault Detection Time}
In this experiment, we aim to measure the time efficiency of our approach in comparison with
that of the baselines. In addition to measuring the average prioritization time for a project in our dataset, we also measured the average time to detect all the faults and the average time to detect one fault in a project (including the prioritization time). As seen in Table~\ref{tab:time}, due to multi-objective optimization strategy, the average prioritization time costs of our approaches for a project (SegTCP and SegTCP*) are much higher than those for the baseline approaches. This is expected due to the optimization algorithms used in our framework (AGE-MOEA over NSGA-II). However, considering the effectiveness of the approaches in fault detection, we can observe that both of our approaches SegTCP and SegTCP* achieve the shortest average time to detect each fault and to detect all the faults in a project.  This demonstrates the time efficiency of our approaches in which despite having longer time for ranking/prioritizing the test cases, the average time to detect faults is smallest, i.e., the effectiveness in fault detection in a time unit is highest.  The total time to detect one fault that our approach saved is from {\bf 15.7\% to 2.46X}. In brief, the time cost for prioritization with multiple objective optimization is worth spending in order to detect faults much earlier.}}

\subsubsection{Segmentation Time}
{\textcolor{custom-blue}{
In accordance with Section 3.1, we reused the DOM-based web page segmentation algorithm from \cite{huynh2023web} to fully automate the web segmentation process. To test one version of a project, the segmentation is conducted once during the initial prioritization run for each test suite, extracting coverage information. Subsequently, this coverage data is stored and utilized for subsequent prioritization runs. The average time required for segmentation per screen is 0.5 seconds, based on our segmentation on 1,300 screens with various complexity. Notably, the time taken to segment the largest test suite, both in terms of the number of test steps and the complexity of a DOM, amounts to 8.6 minutes.
}}

\subsection{Statistical Tests}
\label{sec:stats}
\begin{table}[!htbp]
\centering
\caption{One-sided Wilcoxon Signed-rank Test Results Comparing SegTCP with Other Methods on Performance Metrics.}
\label{tab:mann-whitney}
\begin{tabular}{c|ccc}
\textbf{SegTCP   vs.} & \textbf{APFD} & \textbf{APFDc} & \textbf{FDR} \\ \hline
\textbf{GA} & 0.0183 & 0.0068 & 0.0025 \\
\textbf{GT} & 0.0025 & 0.0005 & 0.0005 \\
\textbf{GA-S} & 0.0005 & 0.021 & 0.0025 \\
\textbf{Terminator} & 0.0093 & 0.0337 & 0.0005 \\
\textbf{ART-F} & 0.0005 & 0.0269 & 0.0005 \\
\textbf{Random} & 0.0005 & 0.021 & 0.0005
\end{tabular}%
\end{table}

We conduct statistical analyses to assess the improvement of segment-based TCP over alternative methods in terms of performance metrics. Combining data from all 11 test suites, we employ the one-sided Wilcoxon Signed-rank test, chosen for its paired samples. We set a p-value threshold of 0.05, wherein p-values less than this threshold indicate that the null hypothesis should be rejected, in other words, the difference in the distributions of our method is significantly higher than the compared method.

At the significance level of 0.05, our method makes significant improvements as compared to other TCP approaches on APFD, APFDc, and FDR. In other words, we can detect faults better than the alternative ones. 

\section{Related Work}
Search-based approaches leverage heuristic searching algorithms to explore the optimal permutation of test cases. There are various implementation for the searching algorithm. Li et al. \cite{li2007search} list out five popular algorithms for this approach: Greedy Algorithm, Additional Greedy, 2-Optimal Algorithms, Hill Climbing and Genetic Algorithms (GenA).
NSGA-II \cite{deb2002fast} is a GenA-based technique applied in solving constrained multi-objective optimization problem. Li et al. \cite{li2013fine} employ and parallelize NSGA-II for better speed in their work. Pradhan et al. \cite{pradhan2018remap} extract rules from past execution data and took them as inputs for their prioritizer. They establish two objectives, namely Fault Detection Capability (FDC) and Test Case Reliance Score (TRS), and incorporate them into NSGA-II to search for optimal solutions. Birchler et al. \cite{birchler2023single} propose an approach for black-box TCP leveraging multi-objective GenAs for testing self-driving cars. In their work, they utilize NSGA-II to explore optimal solutions based on two objectives: the sum of distances between a test $t_i$ and its predecessor $t_{i - 1}$, and the sum of the cost of a test case $t_i$ divided by its position $i$ in the test suite.

Coverage-based strategies \cite{rothermel1999test, rothermel2001prioritizing,hao2015optimal} aim to rank test cases in a manner that maximizes the coverage of target items (e.g. branches, code changes, etc.). The total strategy picks the next test case with the highest absolute coverage, while the additional strategy selects the one with the highest coverage of code units that haven't been covered by the prioritized test cases. 
\cite{mahdieh2020incorporating} enhance coverage-based methods by integrating fault-proneness information, obtained by a neural network into them. \cite{michaels2020combinatorial, michaels2021test} propose 3 new coverage criteria that took into account elements and event sequences based on a hypothesis that combinatorial-based criteria could improve the rate of fault detection. In \cite{michaels2020combinatorial}, the test case that includes sequences occurring only once in the test suite is added into the reduced test suite, and the sequences included are marked as covered. The algorithm in \cite{michaels2021test} greedily search for the next test case that has the highest number of t-way sequences covered. 

Similarity-based approaches try to cover test cases that are distant from the already covered test cases. In \cite{miranda2018fast}, the next test case is the one with the greatest Jaccard similarity to the set of so-far-ordered test cases. ART-based techniques \cite{jiang2009adaptive, zhou2012fault} attempt to spread the test input in the input domain as evenly as possible.

History-based approaches reuse data from previous runs for the current cycle. \cite{huang2012history} utilizes historical data such as the cost of test cases, the faults detected by the test case, and the fault severity of detected faults and input them to GenA to search for the optimal order. \cite{nguyen2021rltcp} combine reinforcement learning and coverage graph to prioritize test cases.

Chen et al. \cite{chen2018optimizing} adopt an XGBoost predictive model to predict which prioritization technique is best suited for current test suite. Yu et al. \cite{yu2019terminator} design a reinforcement learning framework that takes the description of test cases as well as historical data as input to order test cases for UI testing. 





\section{Conclusion}

In conclusion, this paper introduces a multi-objective optimization approach for prioritizing UI test cases, specifically addressing the challenges posed by web page elements. The proposed SegTCP method, utilizing AGE-MOEA search algorithms, outperforms existing TCP approaches, achieving the highest APFD and APFD with Cost (APFDc) at 87.8\% and 79.2\%, respectively. The experiments, conducted on a self-collected dataset comprising 11 diverse test suites, demonstrate the effectiveness of SegTCP in fault detection, element coverage, and minimizing test redundancy. The results also highlight the importance of prioritizing UI test cases, especially in scenarios where test suites are time-consuming, and traditional prioritization methods may not be suitable. The statistical analyses further substantiate the significant improvements offered by SegTCP over alternative methods, emphasizing the effectiveness of WPS in enhancing the efficiency of UI testing. Overall, the contributions of this paper include the introduction of a new optimization approach, the construction of a comprehensive dataset, and empirical evidence showcasing the superiority of SegTCP in UI test case prioritization.

\section*{Acknowledgments}
We thank Katalon inc. for sponsoring this research. Additionally, Tien N. Nguyen is supported in part by the US NSF grant CNS-2120386 and the NSA grant NCAE-C-002-2021, and Vu Nguyen is partially funded by the Vingroup Innovation Foundation (VINIF) under the grant number VINIF.2021.JM01.N2.  

\balance

\bibliographystyle{ACM-Reference-Format}
\bibliography{ref}
\appendix
\end{document}